\documentclass[runningheads]{llncs}
\usepackage{graphicx}
\begin{document}
\title{Educational and Outreach Resource for Astroparticle Physics}
%
%
\author{Y. Kazarina\inst{1} \and
V. Khristyuk\inst{2} \and
A. Kruykov\inst{3} \and
E. Postnikov\inst{3} \and
V. Samoliga\inst{1} \and
A. Shigarov\inst{2} \and
V. Tokareva\inst{4} \and
D. Zhurov\inst{1} 
}
\authorrunning{Y. Kazarina et al.}
%

\institute{Applied Physics Institute of ISU,Irkutsk, Russia \and
Matrosov Institute for System Dynamics and Control Theory, Siberian Branch of Russian Academy of Sciences, Irkutsk, Russia \and
Lomonosov Moscow State University, Skobeltsyn Institute of Nuclear Physics, Moscow, Russia \and
Institute for Nuclear Physics, KIT, Karlsruhe, Germany
}
\maketitle              
\begin{abstract}
The modern astrophysics is moving towards the enlarging of experiments and combining the channels for detecting the highest energy processes in the Universe. To obtain reliable data, the experiments should operate within several decades, which means that the data will be obtained and analyzed by several generations of physicists. Thus, for the stability of the experiments, it is necessary to properly maintain not only the data life cycle, but also the human aspects, for example, attracting, learning and continuity. To this end, an educational and outreach resource has been deployed in the framework of German-Russian Astroparticle Data Life Cycle Initiative (GRADLCI).

\keywords{Astroparticle Physics, TAIGA observatory, Baikal-GVD
neutrino telescope, \texttt{astroparticle.online}, Multi-messenger Astronomy, Deep Learning, CNN, Gamma-Hadron Separation} 
\end{abstract}

\section{Introduction}

The only way to study high-energy processes occurring within and outside the Milky Way is to detect the radiation and ultra-high-energy particles generated by these processes. The flux of ultrahigh-energy cosmic rays, gamma rays and neutrinos interacting with the atmosphere gives rise to cascades of secondary particles. Reaching the ground, these cascades can cover areas of tens of km$^2$, moreover, with an increase in the energy of the initial particle, their flux drops sharply, reaching one particle per year per thousand km$^2$. Thus, over the past few years, experimental astrophysics of ultrahigh energies has been moving towards the enlarging of experiments and combining the channels for detecting high-energy processes named multi-messenger astronomy~\cite{Bartos}. 


The Baikal region is a unique place in Russia since two of the largest setups, investigating three channels of multi-messenger astronomy, are deployed here: the TAIGA gamma observatory~\cite{Budnev}, detecting cosmic rays and gamma rays, and the Baikal-GVD deep underwater telescope, detecting neutrinos~\cite{Avronin}.
Already, the flow of raw experimental data in these setups amounts to several terabytes per day. With the expansion of existing and commissioning of new setups, the data flow will grow many times over, which will lead to a slowdown in the rate of data processing and a decrease in the efficiency of experiments in general. To avoid such a scenario, it is necessary to pay great attention to planning the life cycle of experimental data (from modeling to publishing data in the public domain and archiving data), predicting the volume of data flow and assessing the prospects for using new approaches to data processing to solve new physical problems. Life cycle planning should address such issues as developing new approaches to storing tasks, finding and setting physical tasks that can be solved within the framework of this experiment, assessing the complexity and execution time of tasks related to data analysis. Besides, real-time preliminary data analysis is an important area for development. The presence of an online analysis system will allow one to quickly respond to problems and improve data quality. Also, online analysis will allow TAIGA and Baikal-GVD experiments to be prepared for multi-messenger astronomy and interaction with other setups around the world.

To meet this challenge the new Baikal Multimessenger Lab was established at the Irkutsk State University (ISU) with the support of the Ministry of Education and Science of the Russian Federation this year. The declared missions of the Lab are:
\begin{itemize}
\item Creation of a common framework for experiments in the Baikal region (Baikal-GVD and TAIGA); 
\item Integration of these setups into full-stack multi-messenger astronomy;
\item Creation of a competitive school for astroparticle physics at ISU.
\end{itemize}

 So, the important goal of the Lab is to attract more students to the astroparticle physics and train highly qualified specialists in the field of data processing and analysis for multi-messenger astronomy. The educational and outreach resource \texttt{astroparticle.online} contributes to the achievement of this goal. 
 
 This article is about the resource \texttt{astroparticle.online}~\cite{Kazarina}, \cite{Kazarina2}, its goals and application for Baikal region experiments. In Section 2 there is the description of the resource on the whole and Section 3 is devoted to the interactive part of the resource, developed for gamma-ray astronomy tasks using convolutional neural networks (CNN).

\section {Web Resource \texttt{astroparticle.online}}
\begin{figure}[t]
	\centering
	\includegraphics[width=1\linewidth]{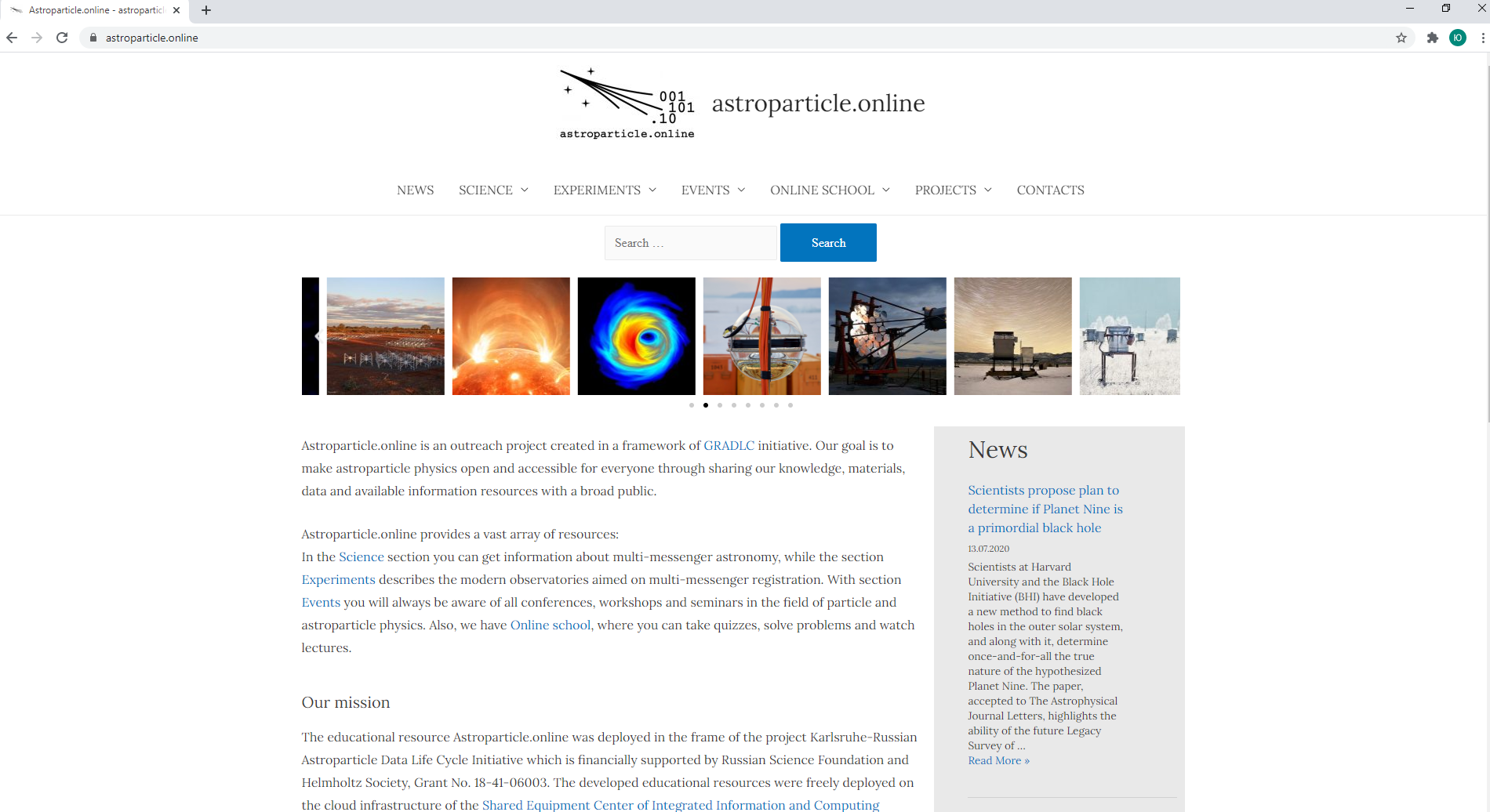}
	\\
	\caption{Screenshot of the first page of the resource \texttt{astroparticle.online}.}
	\label{fig:firstpage}
\end{figure}

The deployment of the \texttt{astroparticle.online} resource  (Fig.~\ref{fig:firstpage}) was started in 2018 in the frame of the German-Russian Astroparticle Data Life Cycle Initiative (GRADLCI)~\cite{Bychkov}. The resource is built on a free and open-source content management system WordPress. The servers of the platform are located at the Matrosov Institute for System Dynamics and Control Theory.
The main target audience –-- students, who are interested in astroparticle physics and would like to collaborate with the Baikal region astroparticle physics experiments.

The resource has several sections: \textit{News} on astronomy and astrophysics (updated weekly), \textit{Science}, \textit{Experiments}, \textit{Projects} dedicate to the Theory of messengers and astroparticle physics experiments and projects, respectively.
These sections aim to attract younger students and schoolchildren and contain text materials, also video materials that are borrowed from other sources with referring to the source, quizzes for better assimilation of the material.

The section \textit{Online School}, the largest one, is mostly original, partly based on the new course in astrophysics launched at the ISU in 2019. It has the following subsections:
\begin{itemize}
\item \textit{Data Analysis};
\item \textit{Lections};
\item \textit{Seminars};
\item \textit{Labs};
\item \textit{Popular Science};
\item \textit{Tasks}.
\end{itemize}


\section{Client on CNN for Gamma/Hadron Separation}
Ground-based gamma-ray astronomy studies very energetic radiation of galactic and extragalactic origin by specially designed telescopes, the so-called Imaging Air Cherenkov Telescopes (IACTs)~~\cite{Weeks}. With this technique, gamma-rays are observed on the ground optically via the Cherenkov light emitted by air-showers of secondary particles when a very-high-energy gamma-ray strikes the atmosphere. Gamma-rays of such energies contribute only a fraction below one per million to the flux of cosmic rays, most of which are protons~\cite{Lorentz}. Nevertheless, being particles without electric charge they can be extrapolated back to their origin, which makes them the best “messengers” of exotic and extreme processes and physical conditions in the Universe. That is why particle identification (gamma-ray discrimination against the cosmic-ray background) is an essential part of data analysis for the IACT technique.

 The subsection \textit{Data Analysis} contains a prototype of the Astroparticle CNN client developed for gamma-ray astronomy tasks as an interactive service for students. This prototype provides access to an on-line analysis of the gamma/hadron separation using convolutional neural networks developed as part of the GRADLCI project~\cite{Postnikov}. The Monte Carlo events of the TAIGA-IACT telescope are used as input for this prototype. The developed convolutional neural network gets the probability of the gamma reconstruction for each event as a result. There is also a possibility to check your skills in gamma/hadron separation using the telescope image.

A shower image is fitted as an ellipse. The ellipse is characterized by its axes and has parameters: length, width, distance and the angular miss-alignment of the major axis. In comparison with hadron showers gamma-ray ones have more elliptic shape, less width and major axis pointed to the source. 
The neural network underlying the prototype takes these features into account. 

\begin{figure}[t]
	\centering
	\includegraphics[width=1\linewidth]{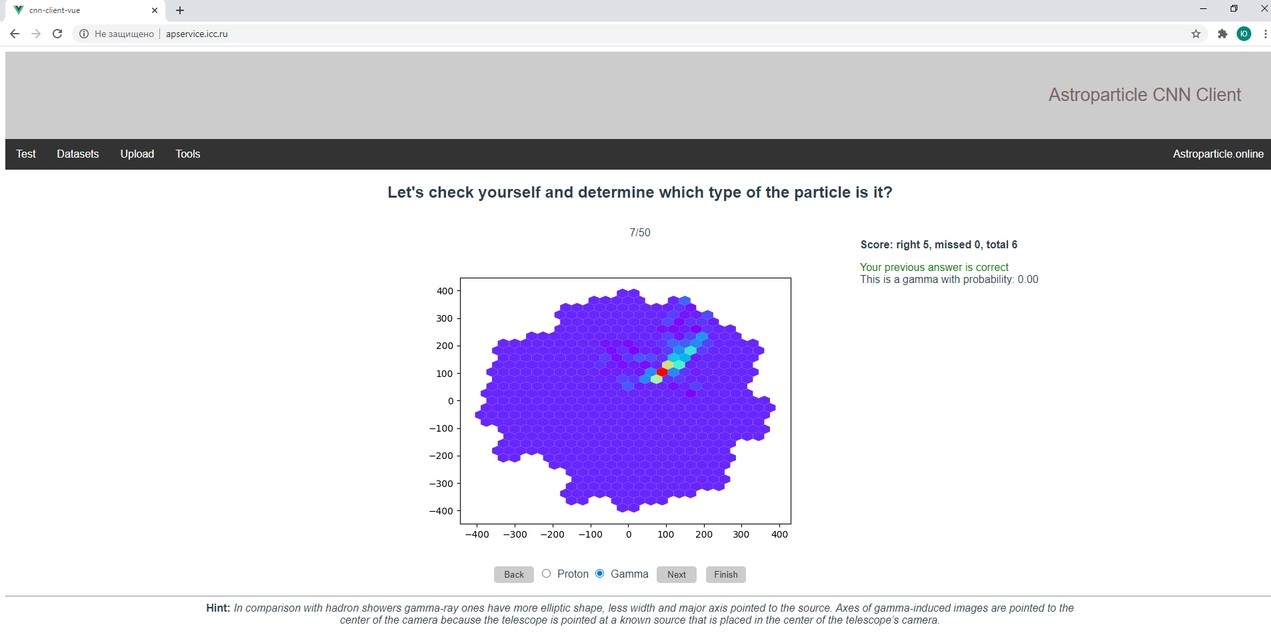}
	\\
	\caption{Screenshot with guessing game on gamma/hadron separation.}
	\label{fig:separation}
\end{figure}

The interactive service contains:
\begin{itemize}
\item Guessing game on gamma/hadron separation (Fig.~\ref{fig:separation});
\item	Instruction how to define gamma-event using telescope image;
\item	Prepared datasets that are ready for downloading;
\item	Application for processing your own dataset;
\item	Tools:
\begin{itemize}
\item	Script for data visualization + instruction; 
\item	Script for converting data files to HDF5 format + instruction.
\end{itemize}
\end{itemize}

The back end of the service is deployed as a microservice in a Docker container. It is written in Python and is based on the Django framework. SQLite is used as the Database Management System, since it contains only information about preloaded datasets. The database includes two tables: a dataset table and an event table. The database structure is defined by the model described using the Django framework. Tables are populated from HDF5 files with data sets during system deployment. Also during deployment of the system the event image files are generated.

\section{Conclusion}

The \texttt{astroparticle.online} resource is intended to become an education and outreach instrument for the new Baikal Multimessenger Lab as well as to advertise the experiments of the Baikal region such as the TAIGA gamma observatory and the Baikal-GVD neutrino telescope. We hope that this resource will attract more students to the astroparticle physics and will allow us to train highly qualified specialists in the field of data processing for astrophysical experiments. This can solve a problem of great importance, namely, to prepare the above experiments for multi-messenger astronomy and for the interaction with other setups around the world.

The resource is filled with materials and tasks in astroparticle physics and its content is regularly updated. The prototype of the Astroparticle CNN client is developed. It is implemented in the resource \texttt{astroparticle.online} as an interactive service for illustrating one of the most complicated challenges in gamma-ray astronomy - the gamma/hadron separation.

\subsubsection{Acknowledgements}
This work was supported by Russian Science Foundation Grant 18-41-06003 (Section 2, 3), by the Helmholtz Society Grant HRSF-0027 and by the Russian Federation Ministry of Science and High Education (project. FZZE-2020-0024). We are grateful to the
members of the GRADLCI for the informational support of our activity.


\begin{thebibliography}{8}

\bibitem{Bartos}
I.~Bartos and M.~Kowalski, \emph{Multimessenger Astronomy}. 2399-2891, IOP Publishing, 2017.
\bibitem{Budnev}
N.~Budnev et al., \emph{TAIGA - a hybrid array for high energy gamma astronomy and
cosmic ray physics}, EPJ Web Conf., vol. \textbf{191}, p. 01007, 2018.
\bibitem{Avronin}
A.~D.~Avrorin et al., \emph{The prototyping/early construction phase of the BAIKAL-GVD project}, Nucl. Instrum. Meth., vol. \textbf{A742}, pp. 82–88, 2014.
\bibitem{Kazarina}
Y.~Kazarina et al., \emph{Towards the Baikal Open Laboratory in Astroparticle Physics}, CEUR Workshop Proceedings, Vol. \textbf{2406}, pp. 1-6, 2019.
\bibitem{Kazarina2}
Y.~Kazarina et al., \emph{Application of HUBzero platform for the educational process in astroparticle physics}, CEUR Workshop Proceedings, Vol. \textbf{2267}, pp. 553-557, 2018.
\bibitem{Bychkov} 
Bychkov, I. et al.\emph{Russian-German Astroparticle Data Life Cycle Initiative}, Data 3(4), 56 (2018).
\bibitem{Weeks}
T.~C.~Weekes et al. \emph{Observation of TeV Gamma Rays from the Crab Nebula Using the Atmospheric Cerenkov Imaging Technique}, Astrophys. J., \textbf{342}, pp.379–395, 1989.
\bibitem{Lorentz}
E.~Lorenz, R.~Wagner \emph{Very-high energy gamma-ray astronomy}, EPJ, H 37, pp. 459–513, 2012.
\bibitem{Postnikov} 
E.Postnikov et al. \emph{Gamma/Hadron Separation in Imaging Air Cherenkov Telescopes Using Deep Learning Libraries TensorFlow and PyTorch}, Journal of Physics: Conference Series, Vol.\textbf{1181}, p. 012048, 2019.
\end{thebibliography}
\end{document}